\documentclass[aps,pra,amssymb,twocolumn,amsmath,superscriptaddress,10pt,tightenlines]{revtex4-1}

\usepackage{comment}

\usepackage{appendix}

\usepackage{amsmath,amssymb} 
\usepackage{mathtools} 
\mathtoolsset{showonlyrefs,showmanualtags}
\usepackage{physics} 
\usepackage{tikz}
\usepackage{bm} 
\usepackage{amsthm} 
\usepackage{ascmac} 
\usepackage{xcolor} 
\usepackage{enumitem} 
\usepackage{graphicx}

\usepackage{here} 
\usepackage[breaklinks,  hyperfootnotes=true, pdfpagelabels, bookmarks, pageanchor]{hyperref}
\hypersetup{%
	colorlinks=true, linktocpage=true, pdfstartpage=1, pdfstartview=FitH, pdfborder={0 0 0},%
	breaklinks=true, pdfpagemode=UseNone, pageanchor=true, pdfpagemode=UseOutlines,%
	plainpages=false, bookmarksnumbered, bookmarksopen=true, bookmarksopenlevel=1,%
	hypertexnames=true, pdfhighlight=/O,
	urlcolor=blue, linkcolor=blue, citecolor=blue, 
}

\usepackage{overpic}

\begin{document}

\title{Exact WKB analysis for adiabatic discrete-level Hamiltonians}

\author{Takayuki Suzuki}
\affiliation{National Institute of Information and Communications Technology, Nukui-Kitamachi 4-2-1, Koganei,
Tokyo 184-8795, Japan}
\author{Eiki Taniguchi}
\affiliation{Department of Physics, Waseda University, Tokyo 169-8555, Japan}
\author{Kaito Iwamura}
\affiliation{Department of Physics, Waseda University, Tokyo 169-8555, Japan}

\begin{abstract}
The dynamics of quantum systems under the adiabatic Hamiltonian has attracted attention not only in quantum control but also in a wide range of fields from condensed matter physics to high-energy physics because of its non-perturbative behavior. Here we analyze the adiabatic dynamics in the two-level systems and the multilevel systems using the exact WKB analysis, which is one of the non-perturbative analysis methods. As a result, we obtain a formula for the transition probability, which is similar to the known formula in the two-level system. Although non-perturbative analysis in the adiabatic limit has rarely been studied for multilevel systems, we show that the same analysis can be applied and also provide a concrete example. The results will serve as a basis for the application of the exact WKB analysis in various fields of physics.
\end{abstract}

\maketitle

\section{Introduction}\label{sec:intro}

Non-perturbative effects in quantum mechanics, which cannot be explained by the perturbative approximation, are becoming important in various areas of physics~\cite{bender1999advanced}. For example, the Dykhne--Davis--Pechukas (DDP) formula~\cite{dykhne1962adiabatic,davis1976nonadiabatic} is known as one of the non-perturbative analysis methods. It is used in a wide range of applications from quantum control~\cite{vitanov1999nonlinear,guerin2002optimization,lehto2012superparabolic,ashhab2022nonlinear} to condensed matter physics~\cite{oka2010dielectric,oka2012nonlinear,kitamura2020nonreciprocal,takayoshi2021nonadiabatic} and high-energy physics~\cite{fukushima2020lefschetz}. In the context of quantum control, the DDP formula describes the dynamics of two-level quantum systems under an adiabatic Hamiltonian, and recently it has been reported that this is a good approximation beyond adiabatic regime for some models~\cite{ashhab2022nonlinear}.

The turning points, the zero points of energy, make an important contribution to the DDP formula. Originally derived for the situation where there is only single turning point, the formula has been extended to the case where there are multiple turning points, which is called the generalized DDP (GDDP) formula~\cite{suominen1992parabolic,vitanov1999nonlinear,guerin2002optimization,lehto2012superparabolic,ashhab2022nonlinear}. Note, however, that there is no rigorous derivation of this formula. Another extension of the DDP formula has recently been derived~\cite{kitamura2020nonreciprocal}, but the relation between these formals is not well studied.

Another non-perturbative method that has recently received attention is the exact Wentzel--Kramers--Brillouin (WKB) analysis~\cite{aoki2002exact,shimada2020numerical,taya2021exact,sueishi2021exact,hashiba2021stokes,enomoto2021exact,taya2021analytical,suzuki2022general,enomoto2022unruh,enomoto2022cosmo}. This allows one to study how the behavior of the asymptotic solutions of the differential equation changes when crossing the Stokes line extending from the turning points. Furthermore, many aspects have already been investigated in the exact WKB analysis, such as the existence of virtual turning points for higher order differential equations~\cite{symposium1994analyse,aoki1998exact,honda2015virtual} and the derivation of the connection matrix when crossing the Stokes line extending from the singular point~\cite{koike2000exact}. For these reasons, we use the exact WKB analysis to study the dynamics of adiabatic quantum systems and confirm the usefulness of the exact WKB analysis as a method for analyzing quantum dynamics. As a result, we obtain results that are extension of previous studies for two-level systems. Non-perturbative analysis of multilevel systems in the adiabatic limit has rarely been studied due to the complexity of the analysis. We also derive a formula that describes the dynamics of adiabatic multilevel systems. The DDP formula extended to multilevel systems has also been studied~\cite{wilkinson2000nonadiabatic}, but our analysis shows that the dynamics of the quantum system is more complex than investigated in this previous study.

The structure of this paper is as follows. In Sec.~\ref{sec:exact_wkb}, an introduction to the exact WKB analysis and some definitions are provided using a simple example. In Sec.~\ref{sec:exact_two}, we apply the exact WKB analysis to the two-level system to obtain the connection formulas. In this section, we also discuss the behavior of the Stokes line using a concrete example. In Sec.~\ref{sec:exact_multi}, we apply the exact WKB analysis to multilevel systems to obtain the connection formulas. Furthermore, we show that the approximation which we derive agrees with the numerical result for a concrete example of the multilevel system. Finally, a short summary is given in Sec.~\ref{sec:conclusion}.

\section{introduction to exact WKB analysis}\label{sec:exact_wkb}

In this chapter, we briefly explain the concept of exact WKB analysis~\cite{dingle1973asymptotic,voros1983return,silverstone1985jwkb,pham1988resurgence,delabaere1997exact,kawai2005algebraic}. For the second order differential equation
\begin{align}
    \left(-\frac{\partial^2}{\partial x^2}+\eta^2 Q(x)\right) \psi(x,\eta)=0,
\end{align}
we consider the WKB solutions, which are the asymptotic solutions to this differential equation. The solution of the differential equation is approximated by a superposition of the WKB solutions. However, when $x$ is regarded as a complex variable, it is well-known that the coefficients of the superposition change discretely on certain lines in the complex plane. This phenomenon is called the Stokes phenomenon and the lines where they change are called the Stokes lines. We call the formula of the change of the coefficients the connection formula.

\begin{figure}[H]
    \centering
    \includegraphics[width=0.6\linewidth]{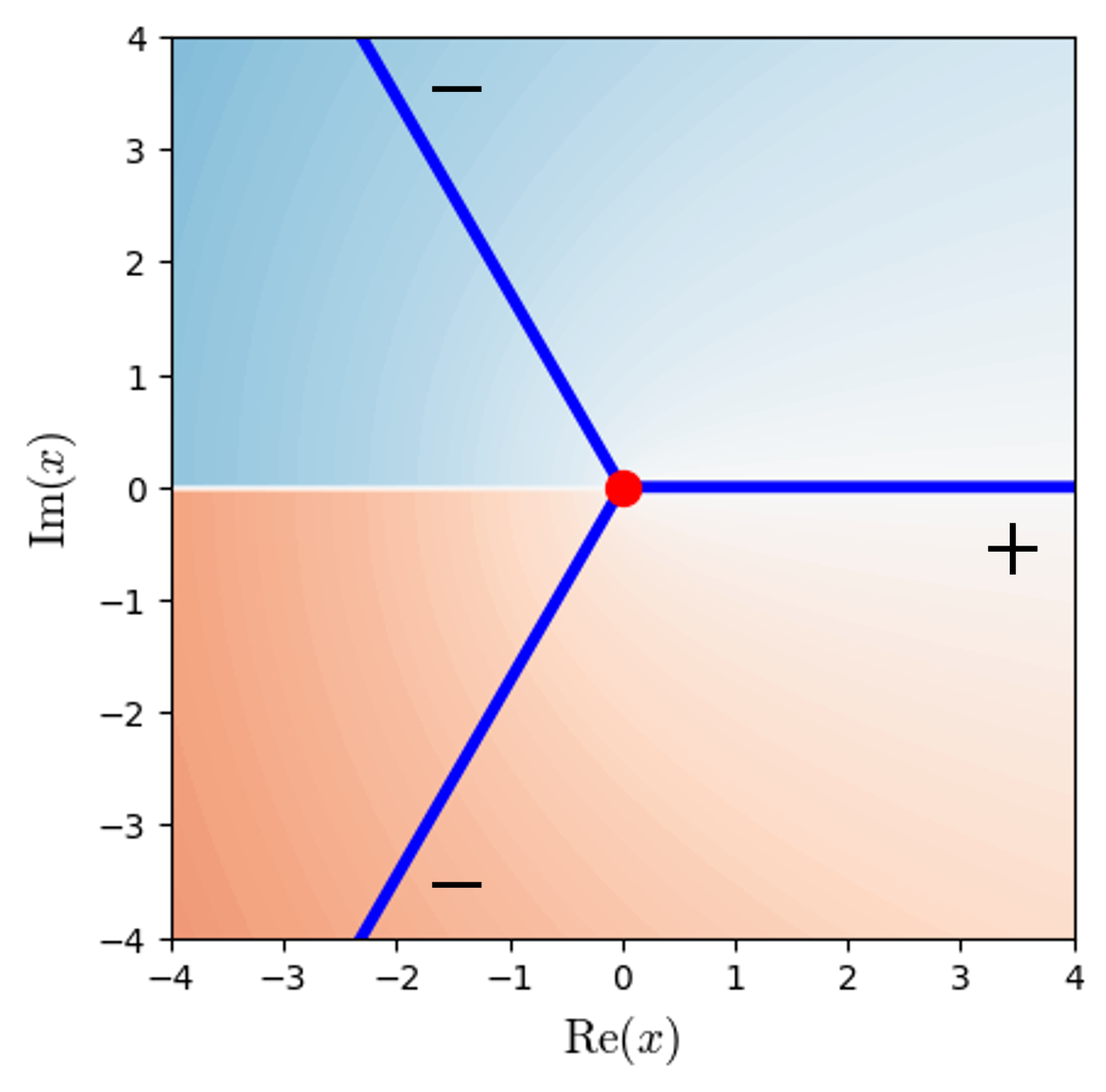}
    \caption{The Stokes lines in the Airy equation ($Q(x)=x$) are plotted. The blue lines represent the Stokes lines; the signs associated with the Stokes lines indicate which WKB solution is dominant on the lines. The background shows $\Im\sqrt{Q(x)}$, and the white line on the negative real axis represents the cut of $\sqrt{Q(x)}$. The red background represents the positive part of $\Im \sqrt{Q(x)}$, and the blue background represents the negative part.}
    \label{fig:stokes_line_airy}
\end{figure}

The Stokes line is defined as the set of $x$ satisfying
\begin{align}
    \operatorname{Im} \int_a^x \sqrt{Q(x')} d x'=0,
\end{align}
where $a$ satisfies $Q(a)=0$ and is called the turning point.
For example, for the Airy equation ($Q(x)=x$), the turning point is $x=0$ and the Stokes line is the set of $x$ satisfying
\begin{align}
    \operatorname{Im} x^{3/2}=0,
\end{align}
which is illustrated in Fig.~\ref{fig:stokes_line_airy}. The WKB solutions 
 of Airy equation are represented as~\cite{bender1999advanced}
\begin{align}
    \psi_+(x,\eta)&=x^{-1/4}e^{\frac{2\eta}{3}x^{3/2}},\\
    \psi_-(x,\eta)&=x^{-1/4}e^{-\frac{2\eta}{3}x^{3/2}}.
\end{align}
The signs associated with each Stokes line in Fig.~\ref{fig:stokes_line_airy} indicate which of the two WKB solutions is dominant when crossing the Stokes line.

When the solutions are connected counterclockwise across the Stokes line where $\psi_+(x,\eta)$ is dominant, the solution varies as in
\begin{widetext}
\begin{align}
    \psi(x,\eta)&\simeq c_+\psi_+(x,\eta)+c_-\psi_-(x,\eta)\\
    &\to c_+\psi_+(x,\eta)+(c_-+ic_+)\psi_-(x,\eta).
\end{align}
On the other hand, when the solutions are connected counterclockwise across the Stokes line where $\psi_-(x,\eta)$ is dominant, the solution varies as in
\begin{align}
    \psi(x,\eta)&\simeq c_+\psi_+(x,\eta)+c_-\psi_-(x,\eta)\\
    &\to (c_++ic_-)\psi_+(x,\eta)+c_-\psi_-(x,\eta).
\end{align}
If the solutions are connected clockwise, the sign in each case is reversed.
\end{widetext}

\section{Exact WKB analysis in the adiabatic limit of the two-level systems}\label{sec:exact_two}

\subsection{Setting}

We introduce some notations for the two-level system that will be discussed in the following sections. First, the adiabatic limit of the Schr\"odinger equation is denoted by 
\begin{align}
    i \frac{\partial}{\partial \tau} \ket{\psi(\tau,\eta)}&=H\qty(\frac{\tau}{\eta}) \ket{\psi(\tau,\eta)}, \label{eq:sch_eq}
\end{align}
where we set $\hbar=1$ and $\eta$ is assumed to be sufficiently large corresponding to the adiabatic dynamics. We note that $\tau$ and $\eta$ have the dimension of time. By introducing a dimensionless quantity $t=\tau/\eta$, Eq.~\eqref{eq:sch_eq} becomes
\begin{align}
    i \frac{\partial}{\partial t} \ket{\psi(t,\eta)}&=\eta H\qty(t) \ket{\psi(t,\eta)} \label{eq:sch_eq_dimless}
\end{align}
and we will consider this type of equation in the following sections.
We note that the Hamiltonian considered in \cite{aoki2002exact,shimada2020numerical}, the term proportional to $\eta$ is diagonal. As a result, they did not consider the adiabatic limit, but a Hamiltonian to which the adiabatic-impulse approximation is applicable~\cite{suzuki2022general}.

The Hamiltonian of the two-level system is generally denoted as 
\begin{align}
    \eta H(t)=\eta d_0(t) I_{2}+\eta \boldsymbol{d}(t) \cdot \boldsymbol{\sigma},
\end{align}
where $\boldsymbol{\sigma}=(\sigma_x,\sigma_y,\sigma_z)$ are the Pauli matrices, $I_n$ is the $n\times n$ identity matrix, and we defined $\boldsymbol{d}(t)=(d_x(t),d_y(t),d_z(t))$. Let $\eta E_{1,2}(t),\ket{E_{1,2}(t)}$ be the eigenvalues and eigenstates of the Hamiltonian $\eta H(t)$ and let $E_1(t)>E_2(t)\ (\forall t\in\mathbb{R})$.
In addition, introducing
\begin{align}
    \tan\theta(t)&=\frac{\sqrt{d_x^2(t)+d_y^2(t)}}{d_z(t)},\\
    \varphi(t)&=\arg \qty(d_x(t)+id_y(t)),
\end{align}
the eigenstates can be expressed as
\begin{align}
    |E_1(t)\rangle&=\left(\begin{array}{c}
    e^{-\frac{i}{2} \varphi(t)} \cos \frac{\theta(t)}{2} \\
    e^{\frac{i}{2} \varphi(t)}\sin \frac{\theta(t)}{2}
    \end{array}\right),\\
    |E_2(t)\rangle&=\left(\begin{array}{c}
    e^{-\frac{i}{2} \varphi(t)} \sin \frac{\theta(t)}{2} \\
    -e^{\frac{i}{2} \varphi(t)}\cos \frac{\theta(t)}{2}
    \end{array}\right).
\end{align}
Expanding the state in terms of these eigenstates
\begin{align}
    \ket{\psi(t,\eta)}&=\sum_{i=1}^2 a_i(t,\eta)\ket{E_i(t)},\label{eq:adi_basis_exp_2}
\end{align}
we get
\begin{widetext}
\begin{align}
    i\frac{\partial}{\partial t}\mqty(a_1(t,\eta)\\ a_2(t,\eta ))
    &=\mqty(\eta \qty(E(t)+E_0(t))+g_{11}(t)&g_{12}(t)\\ g_{21}(t)& -\eta \qty(E(t)-E_0(t))+g_{22}(t))\mqty(a_1(t,\eta)\\ a_2(t,\eta)),\label{eq:adi_sch_2}
\end{align}
\end{widetext}
where we define
\begin{align}
    g_{jk}(t)&=i\bra{\dot E_j(t)}\ket{E_k(t)},\\
    E(t)&=\frac{1}{2}\qty(E_1(t)-E_2(t)),\\
    E_0(t)&=\frac{1}{2}\qty(E_1(t)+E_2(t)).
\end{align}
We represent this equation as
\begin{align}
    i\frac{\partial}{\partial t}\ket{\psi_A(t,\eta)}
    &=H_{A}(t,\eta)\ket{\psi_A(t,\eta)}.\label{eq:adi_sch_2_abbreviation}
\end{align}
In the following, the time $t=t_c$ that satisfies $E(t_c)=0$ is called the turning point. In general, turning points exist in pairs in the upper and lower half planes.

\subsection{Derivation of the connection matrix}

First, we derive the global WKB solutions in \eqref{eq:adi_sch_2_abbreviation}. The WKB solutions can be derived to diagonalize the Hamiltonian formally. To diagonalize $H_A(t,\eta)$, we introduce the matrix
\begin{align}
    R(t,\eta)&=I_2+\sum_{j=1}^\infty \eta^{-j}R_j(t),\\
    R_1(t)&=\mqty(0&\frac{g_{12}(t)}{2E(t)}\\ -\frac{g_{21}(t)}{2E(t)}&0).
\end{align}
Transforming the state as in
\begin{align}
    \ket{\psi_B(t,\eta)}&=R(t,\eta)\ket{\psi_A(t,\eta)},\label{eq:AtoB}
\end{align}
the Eq.~\eqref{eq:adi_sch_2_abbreviation} can be expressed as
\begin{align}
    &i\frac{\partial}{\partial t}\ket{\psi_B(t,\eta)}\\
    &=\mqty(\eta E_1(t)+g_{11}(t)&0\\ 0&\eta E_2(t)+g_{22}(t))\ket{\psi_B(t,\eta)}\\
    &\quad +O(\eta^{-1}).
\end{align}
From this, we get the WKB solutions
\begin{align}
    \ket{\psi_{B,1}(t,t_0,\eta)}&=\mqty(e^{-i\int^t_{t_0} \qty(\eta E_{1}(s)+g_{11}(s)) ds}\\ 0),\\
    \ket{\psi_{B,2}(t,t_0,\eta)}&=\mqty(0\\ e^{-i\int^t_{t_0} \qty(\eta E_2(s)+g_{22}(s))ds}),
\end{align}
where $t_0$ is the reference time. Returning to the original basis with the Eq.~\eqref{eq:AtoB}, we obtain
\begin{align}
    &\ket{\psi_{A,1}(t,t_0,\eta)}\\
    &=
    \mqty(e^{-i\int^t_{t_0} \qty(\eta E_{1}(s)+g_{11}(s)) ds}\\ 0)+O(\eta^{-1}),\label{eq:psi_p}\\
    &\ket{\psi_{A,2}(t,t_0,\eta)}\\
    &=
    \mqty(0\\ e^{-i\int^t_{t_0} \qty(\eta E_{2}(s)+g_{22}(s)) ds})+O(\eta^{-1}).\label{eq:psi_m}
\end{align}
The solutions of the Eq.~\eqref{eq:adi_sch_2_abbreviation} can be approximated by the superposition of these WKB solutions asymptotically. 

Hereafter, we derive connection formulas for these global WKB solutions. To this end, we find the relation between these global WKB solutions and the local WKB solutions corresponding to the Airy equation. The detailed calculations are given in the Appendix~\ref{appsec:connection}.

From here on, it is assumed that the Stokes lines are three lines extending to infinity. From Eq.~\eqref{eq:connection_tc_p} and Eq.~\eqref{eq:connection_tc_m}, the connection matrix of the global WKB solution when crossing counterclockwise the Stokes line where $\ket{\psi_{A,2}(t,t_{c},\eta)}$ is dominant, is expressed as 
\begin{widetext}
\begin{align}
    &\mqty(\ket{\psi_{A,1}(t,t_0,\eta)}\\\ket{\psi_{A,2}(t,t_0,\eta)})\to 
    \mqty(1&0\\i\cot\frac{\theta(t_{c})}{2} e^{-i\int_{t_0}^{t_c} \qty(\eta \qty(E_{2}(s)-E_1(s)) +g_{22}(s)-g_{11}(s))ds} &1)\mqty(\ket{\psi_{A,1}(t,t_0,\eta)}\\\ket{\psi_{A,2}(t,t_0,\eta)}).\label{eq:connect_matrix_2}
\end{align}
On the other hand, the connection matrix of the global WKB solution when crossing counterclockwise the Stokes line where $\ket{\psi_{A,1}(t,t_{c},\eta)}$ is dominant, is expressed as 
\begin{align}
    &\mqty(\ket{\psi_{A,1}(t,t_{0},\eta)}\\\ket{\psi_{A,2}(t,t_{0},\eta)})\to \mqty(1&i\tan\frac{\theta(t_{c})}{2} e^{-i\int_{t_0}^{t_c} \qty(\eta \qty(E_{1}(s)-E_2(s)) +g_{11}(s)-g_{22}(s))ds}\\0&1)\mqty(\ket{\psi_{A,1}(t,t_{0},\eta)}\\\ket{\psi_{A,2}(t,t_{0},\eta)}).\label{eq:connect_matrix_1}
\end{align}
\end{widetext}
We note that this results and the DDP and GDDP formulas agree if the initial state is the ground state. This similarity is discussed in the Appendix.~\ref{appsec:coeff}.

\subsection{Example: nonlinear LZSM model}\label{subsec:example_2level}

We consider the following nonlinear Landau--Zener--St\"uckelbelg--Majorana (LZSM) model~\cite{ashhab2022nonlinear,lehto2012superparabolic,vitanov1999nonlinear}
\begin{align}
    \eta H(t)=\eta\begin{pmatrix}
    vt^n&\Delta^\ast \\
    \Delta&-vt^n
    \end{pmatrix},\label{eq:nLZSM}
\end{align}
where $v>0$ is assumed.

\begin{figure}[H]
    \centering
    \includegraphics[width=0.95\linewidth]{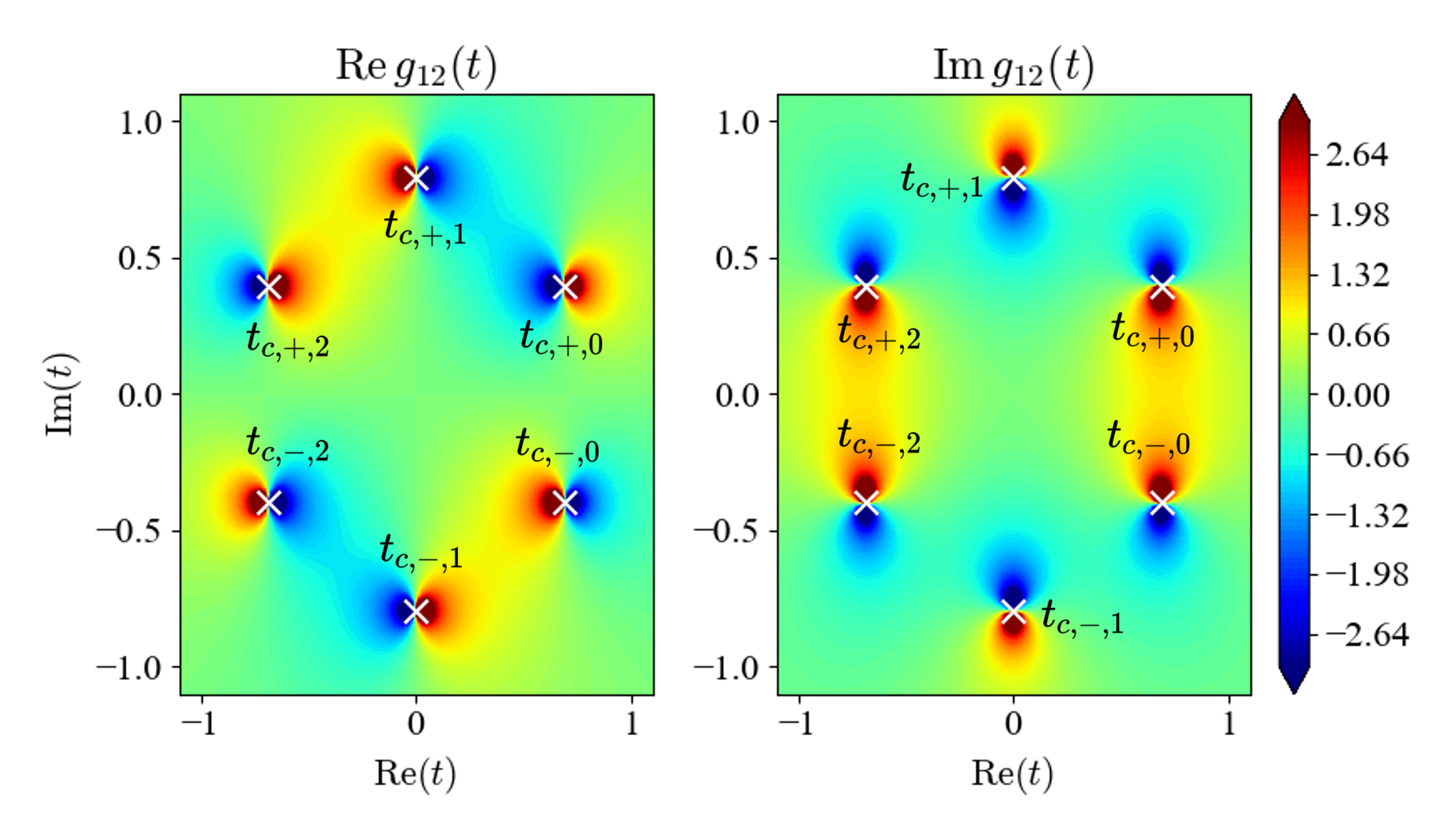}
    \caption{Plots of the real and imaginary parts of $g_{12}(t)$ \eqref{eq:g_comp} in the complex plane. The white crosses represent turning points. It can be seen that $(t-t_c)^{-1}$ diverges near the turning point and that the signs are reversed between adjacent turning points.}
    \label{fig:complex_g_nLZ}
\end{figure}

In this model, the turning points are
\begin{align}
    t_{c,\pm,m}&=e^{\pm i\qty(\frac{\pi}{2n}+ \frac{m}{n}\pi)}\qty(\frac{|\Delta|}{v})^{\frac{1}{n}},\quad m=0,\cdots , n-1
\end{align}
and we get
\begin{align}
    \tan\frac{\theta(t_{c,\pm,m})}{2}
    &=\mp (-1)^m i .\label{eq:tantheta_nLZ}
\end{align}
Also, $g_{12}(t)$ can be expressed as
\begin{align}
    g_{12}(t)
    &=\frac{i}{2}\frac{|\Delta| vnt^{n-1}}{|\Delta|^2+v^2t^{2n}}.\label{eq:g_comp}
\end{align}
This value with $n=3$ is illustrated in Fig.~\ref{fig:complex_g_nLZ}. The sign in \eqref{eq:tantheta_nLZ} changes with $m$, which corresponds to the way $g(t)$ diverges near the turning points, as shown in Fig.~\ref{fig:complex_g_nLZ}.

\begin{figure}[H]
    \centering
    \includegraphics[width=0.95\linewidth]{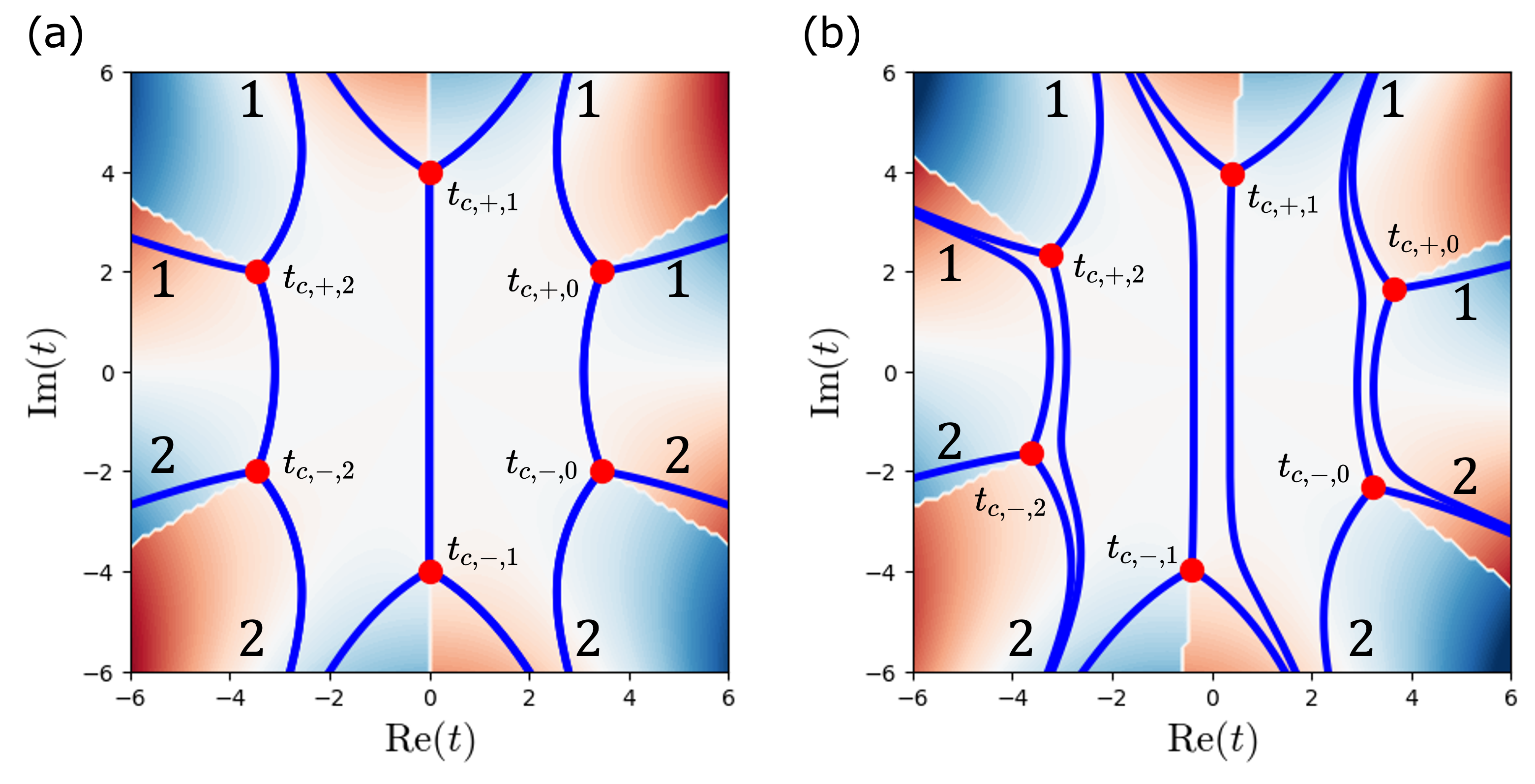}
    \caption{For the Hamiltonian \eqref{eq:nLZSM} with $n=3$, (a) the Stokes line and (b) the Stokes line when a small perturbation $v\to v(1+i\epsilon)$ is added to resolve the degeneracy are plotted. The red dots represent turning points and the blue lines represent the Stokes lines; the numbers associated with the Stokes lines represent the dominant WKB solutions on the lines. After a small perturbation, the degeneracy of the Stokes lines is resolved, and the three Stokes lines extending from the turning point can be seen to extend to infinity without conflicting with other turning points or Stokes lines. The background shows $\Im(E_1(t)-E_2(t))$, and the white lines represent the cut of $E_1(t)-E_2(t)$. The red background represents the positive part of $\Im(E_1(t)-E_2(t))$, and the blue background represents the negative part.}
    \label{fig:stokes_3LZ}
\end{figure}

The Stokes line for this model is depicted in Fig.~\ref{fig:stokes_3LZ}. We note that a small perturbation $v\to v(1+i\epsilon)$ is introduced so that the Stokes lines emerging from the turning points extend to infinity. Although the sign of $\epsilon$ makes the Stokes graph different, this does not affect the transition probability. The treatment of the difference has been discussed in the resurgence theory~\cite{dillinger1993resurgence,delabaere1997exact,shen2008observations}.

\begin{widetext}
Using the formulas~\eqref{eq:connect_matrix_2} and \eqref{eq:connect_matrix_1}, we get the connection formula from $t_0=-\infty$ to $t=\infty$ as
\begin{align}
    \mqty(\ket{\psi_{A,1}(t,t_0,\eta)}\\\ket{\psi_{A,2}(t,t_0,\eta)})\to \mqty(1&\sum_{k=0}^{n-1}(-1)^ke^{-i\eta\int_{t_0}^{t_{c,-,k}} \qty(E_{1}(s)-E_2(s))ds}\\
    -\sum_{k=0}^{n-1}(-1)^ke^{-i\eta\int_{t_0}^{t_{c,+,k}} \qty(E_{2}(s)-E_1(s))ds}&1)\mqty(\ket{\psi_{A,1}(t,t_0,\eta)}\\\ket{\psi_{A,2}(t,t_0,\eta)}).
\end{align}
The transition probability from the ground state at $t=-\infty$ to the excited state at $t=\infty$ in the adiabatic limit is consistent with the result in~\cite{vitanov1999nonlinear,lehto2012superparabolic}. We note that the initial state is not assumed to be the ground state in the formula which we derived, unlike in the DDP formula and the GDDP formula (Appendix.~\ref{appsec:coeff}.).

\end{widetext}

\section{Exact WKB analysis in the adiabatic limit of the multilevel systems}\label{sec:exact_multi}

\subsection{Setting}
In this section, we consider the following $N$-level system 
\begin{align}
    i\frac{\partial}{\partial t}\ket{\psi(t,\eta)}=\eta H(t)\ket{\psi(t,\eta)}.
\end{align}
This setting can be applied, for example, to a system of a driven spin and single-mode boson~\cite{werther2019davydov,sun2012photon,lidal2020generation,wang2021schrödinger,zheng2021photon,neilinger2016landau,bonifacio2020landau}.
The eigenvalues and eigenstates of the Hamiltonian $\eta H(t)$ are denoted as $\eta E_i(t),\ket{E_i(t)}$, respectively, where $\eta>0$ and $E_i(t)>E_{i+1}(t)$. When the state $\ket{\psi(t,\eta)}$ is expanded as
\begin{align}
    \ket{\psi(t,\eta)}=\sum_{i=1}^N a_i(t,\eta)\ket{E_i(t)}.
\end{align}
$a_i(t,\eta)$ satisfies the differential equation
\begin{widetext}
\begin{align}
    i\frac{\partial}{\partial t}\mqty(a_1(t,\eta)\\\vdots\\a_N(t,\eta))&=
    \mqty(\eta E_1(t)+g_{11}(t)&g_{12}(t)&\cdots &g_{1N}(t)\\
    \vdots&\vdots&\cdots&\vdots\\
    g_{N1}(t)&g_{N2}(t)&\cdots &\eta E_N(t)+g_{NN}(t))
    \mqty(a_1(t,\eta)\\\vdots\\a_N(t,\eta)),
\label{eq:multilevel_dif_eq_matrix}
\end{align}
where we define
\begin{align}
    g_{jk}(t)&=i\bra{\dot E_j(t)}\ket{E_k(t)}.
\end{align}

\subsection{The global WKB solution}
Here, we find the WKB solutions of Eq.~\eqref{eq:multilevel_dif_eq_matrix}. To this end, using the matrix
\begin{align}
    R(t,\eta)=I_N+\sum_{j=1}^\infty \eta^{-j}R_j(t),\quad R_1(t)=\mqty(0&\frac{g_{12}(t)}{E_1(t)-E_2(t)}&\cdots &\frac{g_{1N}(t)}{E_1(t)-E_N(t)}\\ 
    \frac{g_{12}^\ast(t)}{E_2(t)-E_1(t)}&0&\dots &\frac{g_{2N}^\ast(t)}{E_2(t)-E_N(t)}\\
    \vdots&\vdots&\ddots&\vdots\\
    \frac{g_{1N}^\ast(t)}{E_N(t)-E_1(t)}&\frac{g_{2N}^\ast(t)}{E_N(t)-E_2(t)}&\dots &0),
\end{align}
we perform the formal diagonalization of the Hamiltonian in the adiabatic basis.
To transform the basis as in
\begin{align}
    \ket{\psi_B(t,\eta)}&=R(t,\eta)\ket{\psi_A(t,\eta)},
\end{align}
the differential equation is represented as 
\begin{align}
    i\frac{\partial}{\partial t}\ket{\psi_B(t,\eta)}
    &=\mqty(\eta E_1(t)+g_{11}(t)&0&\dots &0\\ 0&\eta E_2(t)+g_{22}(t)&\dots &0\\
    \vdots&\vdots&\dots&\vdots\\
    0&0&\dots&\eta E_N(t)+g_{NN}(t)
    )\ket{\psi_B(t,\eta)}+O(\eta^{-1}).
\end{align}
\end{widetext}
The WKB solutions are expressed as
\begin{align}
    \ket{\psi_{B,j}(t,t_0,\eta)}=e^{-i\int^t_{t_0} \qty(\eta E_j(s)+g_{jj}(s))ds}\mathbf{e}_j,&\\
     (j=1,\cdots,N),&
\end{align}
where $\mathbf{e}_j$ is a unit vector with only the $j$-th component $1$ and the rest $0$.
In the original basis, we obtain
\begin{align}
    \ket{\psi_{A,j}(t,t_0,\eta)}&=R^{-1}(t,\eta)\ket{\psi_{B,1}(t,\eta)}\\
    &=e^{-i\int^t_{t_0} \qty(\eta E_j(s)+g_{jj}(s))ds}\mathbf{e}_j+O(\eta^{-1}),\\
    &\quad \quad \quad (j=1,\cdots,N).
\end{align}

\subsection{The construction of the corresponding two-level Hamiltonian}\label{subsec:multi_to_two_ham}
To reduce the discussion of the multilevel system to that of a two-level system, we focus on the behavior near the turning point $t=t_{c,12}$, where $E_1(t_{c,12})=E_2(t_{c,12})$. We assume that there is no time when more than three energies are degenerate, and three Stokes lines extend from the turning point to infinity. We consider the following transformation
\begin{align}
    \ket{\psi_A(t,\eta)}&=\qty(I_N+\sum_{j=1}^\infty \eta^{-j}V_j(t))\ket{\tilde{\psi}_A(t,\eta)},\\
    V_1(t)&=\left(\begin{array}{ccccc}
    0 & 0 & \frac{g_{13}}{E_3-E_1} & \frac{g_{14}}{E_4-E_1} & \cdots \\
    0 & 0 & \frac{g_{23}}{E_3-E_2} & \frac{g_{24}}{E_4-E_2} & \cdots \\
    -\frac{g^\ast_{13}}{E_3-E_1} & -\frac{g^\ast_{23}}{E_3-E_2} & 0 & 0 & \cdots \\
    -\frac{g^\ast_{14}}{E_4-E_1} & -\frac{g^\ast_{24}}{E_4-E_2} & 0 & 0 & \cdots \\
    \vdots & \vdots & \vdots & \vdots & \ddots
    \end{array}\right).
\end{align}
After the transformation, the differential equation for $a_1(t,\eta)$ and $a_2(t,\eta)$ is expressed as
\begin{align}
    &i\frac{\partial }{\partial t}\mqty(a_1(t,\eta)\\ a_2(t,\eta))\\ &=\mqty(\eta E_1(t)+g_{11}(t)&g_{12}(t)\\ g_{21}(t)& \eta E_2(t)+g_{22}(t))\mqty(a_1(t,\eta)\\ a_2(t,\eta))\\
    &\quad +O(\eta^{-1})
\label{eq:multilevel_dif_eq}
\end{align}
because $a_1(t,\eta)$ and $a_2(t,\eta)$ are decoupled from the other variables. We denote this equation as
\begin{align}
    i\frac{\partial }{\partial t}\ket{\tilde{\psi}_A^{(12)}(t,\eta)}&=H_{A}^{(12)}(t,\eta)\ket{\tilde{\psi}_A^{(12)}(t,\eta)}+O(\eta^{-1}).
\end{align}
The previous study~\cite{wilkinson2000nonadiabatic} has already presented the general form of this equation up to higher orders. If there is no singularity in higher orders at $t=t_{c,12}$, the Stokes phenomenon can be adequately described by the lower orders. In other words, if the Stokes lines in this space are shaped like the Airy equation near each turning point, the transition probabilities can be obtained by looking specifically at how the lower orders are determined, as described next.

Now, we consider the Hamiltonian $\tilde H_C^{(12)}(t,\eta)$ in the original basis corresponding to the Hamiltonian $H_A^{(12)}(t,\eta)$ in the adiabatic basis. We assume that the corresponding Hamiltonian can be denoted as
\begin{align}
    \tilde H_C^{(12)}(t,\eta)&=(\eta\boldsymbol{d}_{-1}(t)+\boldsymbol{d}_{0}(t))\cdot \boldsymbol{\sigma}\\
    &\quad +(\eta \omega_{-1}(t)+\omega_0(t))I_2+O(\eta^{-1}).
\end{align}
Then, the eigenstates are
\begin{align}
    |E_1^{(12)}(t)\rangle&=\left(\begin{array}{c}
    e^{-\frac{i}{2} \varphi(t)} \cos \frac{\theta(t)}{2} \\
    e^{\frac{i}{2} \varphi(t)}\sin \frac{\theta(t)}{2}
    \end{array}\right)+O(\eta^{-1}),\\
    |E_2^{(12)}(t)\rangle&=\left(\begin{array}{c}
    e^{-\frac{i}{2} \varphi(t)}\sin \frac{\theta(t)}{2} \\
    -e^{\frac{i}{2} \varphi(t)}\cos \frac{\theta(t)}{2}
    \end{array}\right)+O(\eta^{-1}).
\end{align}
From the Schr\"odinger equation
\begin{align}
    i\frac{\partial}{\partial t}\ket{\tilde\psi^{(12)}(t,\eta)}=\tilde H_C^{(12)}(t,\eta)\ket{\tilde\psi^{(12)}(t,\eta)}
\end{align}
and the expansion of the state
\begin{align}
    \ket{\tilde \psi^{(12)}(t,\eta)}&=\sum_{i=1}^2a_i(t,\eta)\ket{E^{(12)}_i(t)},
\end{align}
we get the Hamiltonian in the adiabatic basis.
By comparing the result with \eqref{eq:multilevel_dif_eq}, we get
\begin{align}
    \sqrt{\boldsymbol{d}_{-1}^2(t)}&=\frac{1}{2}\qty(E_1(t)-E_2(t)),\\
    \omega_{-1}(t)&=\frac{1}{2}\qty(E_1(t)+E_2(t)),\\
    \theta(t)-\theta(t_{c,12})&=-2\int_{t_{c,12}}^t\Im g_{12}(s)ds\label{eq:def_theta}\\
    \varphi(t)-\varphi(t_{c,12})&=-2\int_{t_{c,12}}^t\csc\theta(s)\Re g_{12}(s)ds\\
    \boldsymbol{d}_{-1}(t)\cdot \boldsymbol{d}_{0}(t)&=\frac{\sqrt{\boldsymbol{d}_{-1}^2(t)}}{2}\qty(g_{11}(t)-g_{22}(t)+\dot\varphi(t)\cos\theta(t))\\
    \omega_0(t)&=\frac{1}{2}\qty(g_{11}(t)+g_{22}(t)).
\end{align}
In this way, $\tilde H_C^{(12)}(t,\eta)$ is derived from $H_A^{(12)}(t,\eta)$.

Using the matrix
\begin{align}
    S(t):=\mqty(e^{-\frac{i}{2}\varphi(t)}\cos\frac{\theta(t)}{2}& e^{\frac{i}{2}\varphi(t)}\sin \frac{\theta(t)}{2}\\ e^{-\frac{i}{2}\varphi(t)}\sin\frac{\theta(t)}{2}&-e^{\frac{i}{2}\varphi(t)}\cos \frac{\theta(t)}{2}),
\end{align}
we get
\begin{align}
    \ket{\tilde\psi^{(12)}(t,\eta)}&=\sum_{i=1}^2 a_i(t,\eta)\ket{E^{(12)}_i(t)}\\
    &=S(t)\ket{\tilde \psi^{(12)}_A(t,\eta)},
\end{align}
and we see that the adiabatic and standard bases are related by the matrix $S(t)$. Therefore, we introduce
\begin{align}
    \ket{ \psi_C(t,\eta)}:=e^{-i\eta\int_{t_{c,12}}^t  \omega_{-1}(s)ds}S(t)\ket{\tilde \psi_A^{(12)}(t,\eta)},
\end{align}
and the equation satisfied by $\ket{\psi_C(t,\eta)}$ is
\begin{align}
    i\frac{d}{dt}\ket{ \psi_C(t,\eta)}&=\qty(\eta\boldsymbol{d}_{-1}(t)\cdot\boldsymbol{\sigma}+O(\eta^{0}))\ket{\psi_C(t,\eta)}.
\end{align}
In this way, the Hamiltonian of the two-level system in the standard basis was derived; the method of attributing the two-level system to the Airy equation is discussed in the Appendix~\ref{appsec:connection}.

\subsection{Connection of local and global WKB solutions}\label{subsec:connection_multi}
The local WKB solutions of the Airy equation can be expressed as \eqref{eq:phi_p} and \eqref{eq:phi_m}. In the two-level system, the connection of local and global WKB solutions is given by \eqref{eq:connect_p} and \eqref{eq:connect_m} with $\omega(t)=\omega_{-1}(t)$. Applying the results to the multilevel systems, we need to consider the effect of $V(t)$. In this way, we obtain the following connection formula. The connection matrix of the global WKB solution when crossing counterclockwise the Stokes line where $\ket{\psi_{A,2}(t,t_{c,12},\eta)}$ is dominant is expressed as 
\begin{widetext}
\begin{align}
    \mqty(\ket{\psi_{A,1}(t,t_{0},\eta)}\\\ket{\psi_{A,2}(t,t_{0},\eta)}\\ \vdots\\\ket{\psi_{A,N}(t,t_{0},\eta)})&\to M^{[2>1]}(t_{c,12})
    \mqty(\ket{\psi_{A,1}(t,t_{0},\eta)}\\\ket{\psi_{A,2}(t,t_{0},\eta)}\\ \vdots\\\ket{\psi_{A,N}(t,t_{0},\eta)}),\\
    M^{[2>1]}(t_{c,12})&:=\left(\begin{array}{@{}cc@{}}
    \begin{matrix}
    1 & 0 \\
    i\cot\frac{\theta(t_{c,12})}{2} e^{-i\int_{t_0}^{t_{c,12}} \qty(\eta \qty(E_{2}(s)-E_1(s) )+g_{22}(s)-g_{11}(s))ds}& 1
    \end{matrix}
    & \mbox{\Large $0$} \vspace{3mm}\\\vspace{3mm}
    {\mbox{\Large $0$}} &
    \begin{matrix}
    {\mbox{\Large{$I_{N-2}$}}}
    \end{matrix}
    \end{array}\right).
\end{align}
On the other hand, the connection matrix of the global WKB solution when crossing counterclockwise the Stokes line where $\ket{\psi_{A,1}(t,t_{c},\eta)}$ is dominant is expressed as 
\begin{align}
    \mqty(\ket{\psi_{A,1}(t,t_{0},\eta)}\\\ket{\psi_{A,2}(t,t_{0},\eta)}\\ \vdots\\\ket{\psi_{A,N}(t,t_{0},\eta)})&\to M^{[1>2]}(t_{c,12})
    \mqty(\ket{\psi_{A,1}(t,t_{0},\eta)}\\\ket{\psi_{A,2}(t,t_{0},\eta)}\\ \vdots\\\ket{\psi_{A,N}(t,t_{0},\eta)}),\\
    M^{[1>2]}(t_{c,12})&:=\left(\begin{array}{@{}cc@{}}
    \begin{matrix}
    1 & i\tan\frac{\theta(t_{c,12})}{2}e^{-i\int_{t_0}^{t_{c,12}} \qty(\eta \qty(E_{1}(s)-E_2(s))+g_{11}(s)-g_{22}(s))ds} \\
    0 & 1
    \end{matrix}
    & \mbox{\Large $0$} \vspace{3mm}\\\vspace{3mm}
    {\mbox{\Large $0$}} &
    \begin{matrix}
    {\mbox{\Large{$I_{N-2}$}}}
    \end{matrix}
    \end{array}\right).
\end{align}
\end{widetext}
The $\theta(t)$ in the matrices can be calculated with \eqref{eq:def_theta}. We assume, however, that the $\theta(t_{c,12})$ depends on the behavior of $g_{12}(t)$ near the turning point $t_{c,12}$ which diverges like the two-level systems. We will show in the next example that this assumption is valid.

\subsection{Example}
\begin{figure}[H]
    \centering
    \includegraphics[width=0.7\linewidth]{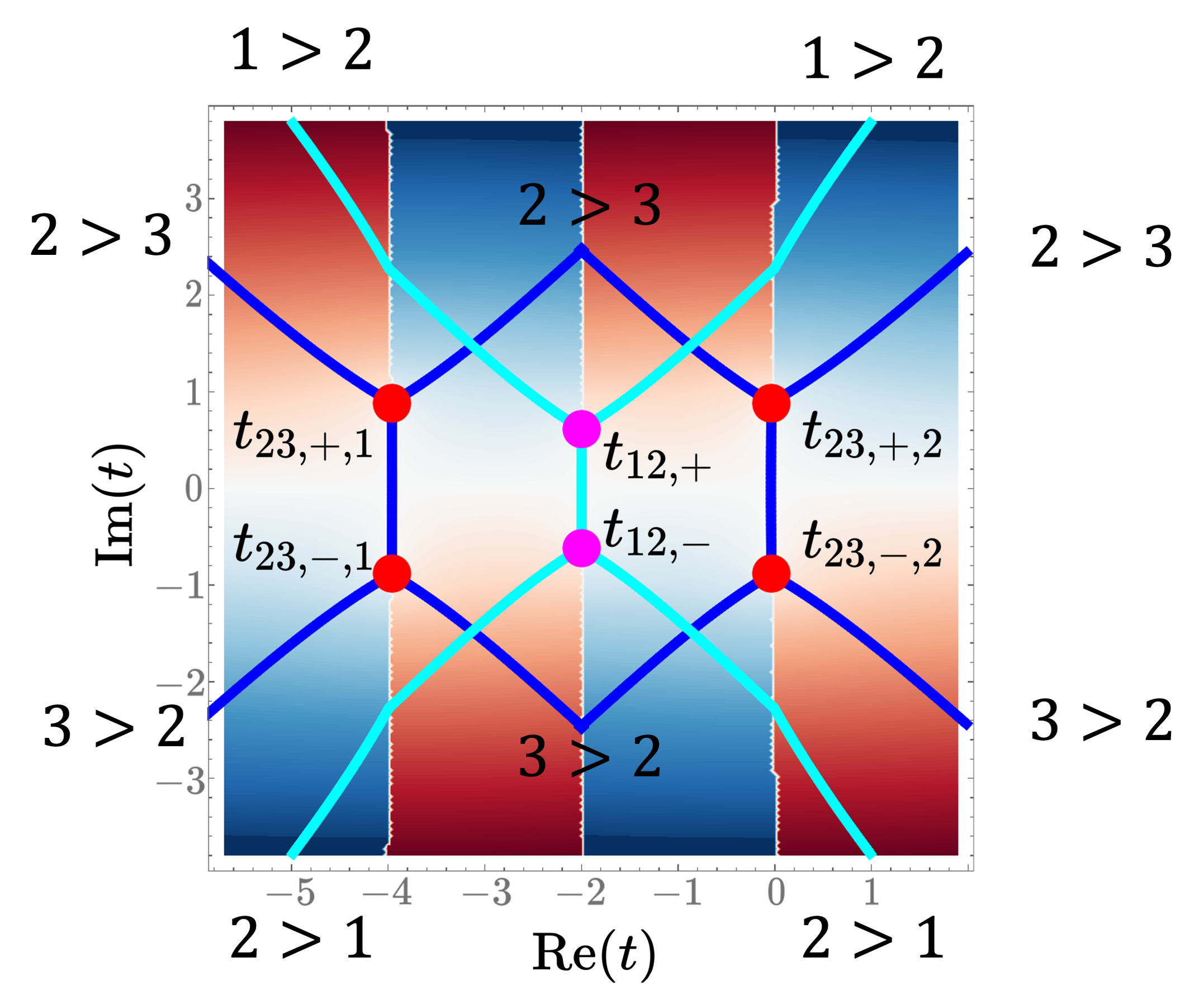}
    \caption{The Stokes line for the Hamiltonian~\eqref{eq:hamiltonian_3level} is plotted. The red dots represent the turning points for $E_2(t)-E_3(t)$ and the pink dots represent the turning points for $E_1(t)-E_2(t)$. The blue lines represent the Stokes lines for $E_2(t)-E_3(t)$ and the cyan lines represent the Stokes lines for $E_1(t)-E_2(t)$. The inequalities associated with the Stokes lines are interpreted as follows. The larger (smaller) number in the inequality represents the dominant (subdominant) WKB solution on the line. The background shows $\Im(E_2(t)-E_3(t))$, and the white lines represent the cut of $E_2(t)-E_3(t)$. The red background represents for the positive part of $\Im(E_2(t)-E_3(t))$ and the blue back ground represents for the negative part of $\Im(E_2(t)-E_3(t))$. We note that, unlike the example of the two-level system, the cuts also come from $t_{12,\pm}$, which are not the turning points of $E_2(t)-E_3(t)$.}
    \label{fig:stokes_3level}
\end{figure}

\begin{figure}[H]
    \centering
    \includegraphics[width=0.95\linewidth]{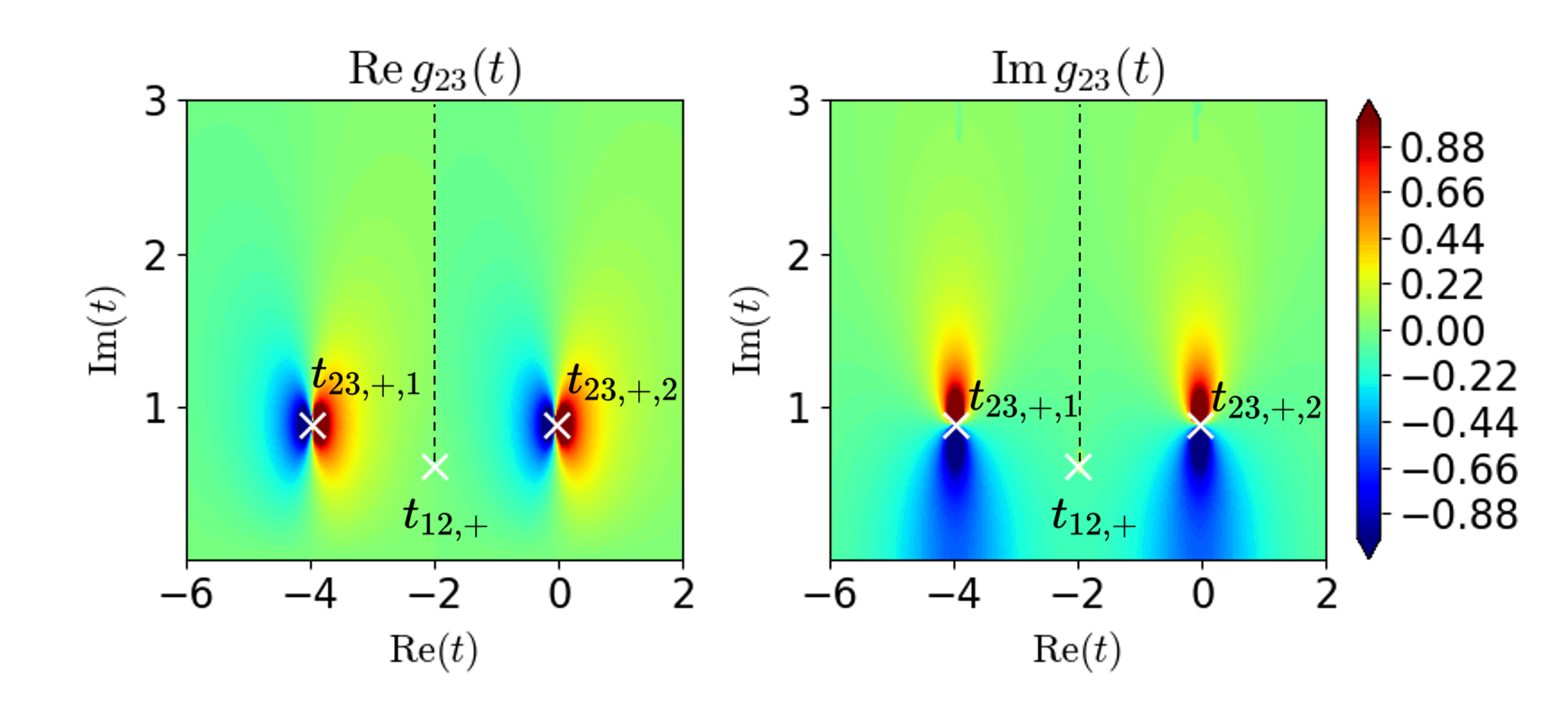}
    \includegraphics[width=0.95\linewidth]{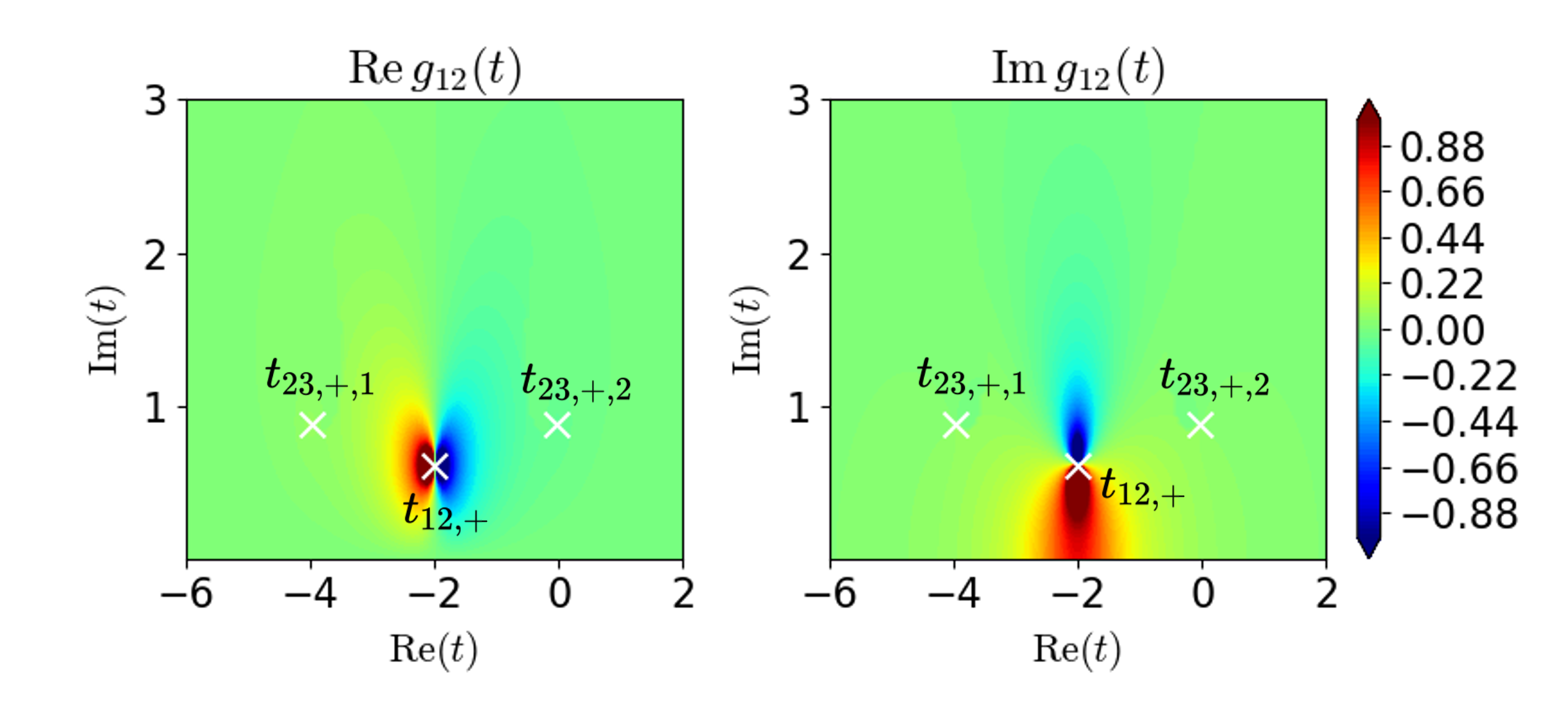}
    \caption{Plots of the real and imaginary parts of $g_{23}(t)$ (top) and $g_{12}(t)$ (bottom) in the upper half plane. The white crosses represent turning points. It can be seen that $(t-t_c)^{-1}$ diverges near the turning point but that the signs are not reversed between adjacent turning points in $g_{23}(t)$ because of the cut extending from $t_{12,+}$ which is represented by the dashed line.}
    \label{fig:g_3level}
\end{figure}

Here, we consider the following 3-level LZSM model~\cite{carroll1986transition,kiselev20133,band2019three}:
\begin{align}
    H(t,\eta)=\eta\mqty(v_1 t&\Delta_{12}&\Delta_{13}\\\Delta_{12}&v_2t+a&\Delta_{23}\\\Delta_{13}&\Delta_{23}&0).\label{eq:hamiltonian_3level}
\end{align}
This model of a time-dependent three-level system describes, for example, the system of nitrogen vacancy centers in diamond~\cite{doherty2013nitrogen,ajisaka2016decoherence,band2019three}. Let $t_{23,+,1},t_{23,+,2},t_{12,+}$ denote the turning points in the upper half plane (see Fig.~\ref{fig:stokes_3level}). We note that $t_{ij}$ corresponds to the zero point of $E_i(t)-E_j(t)$, respectively.
The Stokes lines are shown in Fig.~\ref{fig:stokes_3level}. We note that $E_2(t)-E_3(t)$ has a cut with $t_{12,+}$ as the branch point, and the Stokes lines extending from $t_{23,+,1},t_{23,+,2}$ go through this cut and extend to another Riemann sheet. There are also some Stokes lines that emerge from the virtual turning points which are intersections of the Stokes lines~\cite{symposium1994analyse,aoki1998exact,honda2015virtual}, but these are not shown in Fig.~\ref{fig:stokes_3level} because they are not affected in the present parameter domain.
The $g_{ij}(t)$ in the upper half plane are shown in Fig.~\ref{fig:g_3level}. It can be seen that $g_{ij}(t)\propto(t-t_c)^{-1}$ holds near the turning point.
Note, however, that as can be seen from the behavior of $g_{23}(t)$, there exists a cut with $t_{12,+}$ as the branch point, so that the same singularity exists near both turning points, which is different from the example of the two-level system in Sec.~\ref{subsec:example_2level}.

\begin{figure}[H]
    \centering
    \includegraphics[width=0.85\linewidth]{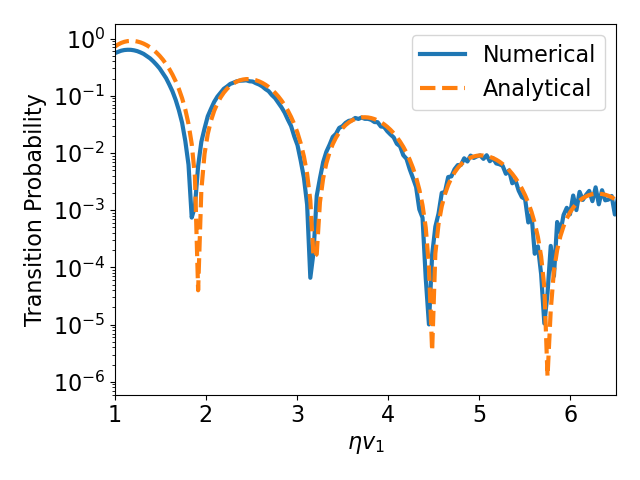}
    \caption{Plot of the transition probability from $\ket{E_3(-\infty)}$ at time $t=-\infty$ to $\ket{E_2(\infty)}$ at time $t=\infty$. The solid line represents the numerical results and the dashed line represents the analytical approximation. It can be seen that they are in good agreement. We used the solver for numerical calculation implemented in the Python library QuTiP~\cite{Johansson2012-du,Johansson2013-gw}.}
    \label{fig:prob_3level}
\end{figure}

As in the example of two-level system, by resolving the degeneracy of the Stokes lines with small perturbations and applying the connection formula, the connection matrix when all the Stokes lines are crossed is 
\begin{widetext}
\begin{align}
    \mqty(\ket{\psi_{A,1}(t,t_0,\eta)}\\\ket{\psi_{A,2}(t,t_0,\eta)}\\\ket{\psi_{A,3}(t,t_0,\eta)})&\to M^{[2>3]}(t_{23,+,2}) \qty(M^{[3>2]}(t_{23,-,2}) )^{-1}M^{[1>2]}(t_{12,+})\\
    &\quad \times\qty(M^{[2>1]}(t_{12,-}))^{-1}M^{[2>3]}(t_{23,+,1}) \qty(M^{[3>2]}(t_{23,-,1}) )^{-1}\mqty(\ket{\psi_{A,1}(t,t_{12,+,1},\eta)}\\\ket{\psi_{A,2}(t,t_{12,+,1},\eta)}\\\ket{\psi_{A,3}(t,t_{12,+,1},\eta)}),
\end{align}
where the energy integral contained in the matrix $M(t)$ is assumed to take an integral path avoiding the cut.
Here we introduced
\begin{align}
    M^{[i>j]}(t)&:=I+i\cot\frac{\theta(t)}{2} e^{-i\int_{t_0}^{t} \eta \qty(E_{i}(s)-E_j(s) )ds}Q_{ij}\quad i>j,\\
    M^{[i>j]}(t)&:=I+i\tan\frac{\theta(t)}{2} e^{-i\int_{t_0}^{t} \eta \qty(E_{i}(s)-E_j(s) )ds}Q_{ij}\quad i<j,
\end{align}
where $Q_{ij}$ is the matrix only $(i,j)$ element of which is $1$ and the others are $0$.
\end{widetext}
Figure \ref{fig:prob_3level} compares this analytical approximation with the numerical calculation. It can be seen that the analytical approximation can indeed be approximated with good enough accuracy in the region where $\eta v_1$ is sufficiently large. 
Note, however, that in the region $\Delta_{23}\simeq 0$, the result of the numerical calculation and the analytical approximation do not match even if $\eta v_1$ is of similar magnitude. The reason for this is that $\eta$ is not large enough. In fact, in the region of $\Delta_{23}\simeq 0$, $\Im t_{12,+}$ becomes small, so that adiabatic evolution does not occur unless $\eta v_1$ is sufficiently large. If the ground state is the initial state and only the lower energy levels are adiabatic, i.e. $\Im t_{23,+,i}$ is large, the approximation fails when $\Im t_{12,+}$ is small.

Finally, note the relation to the previous study~\cite{wilkinson2000nonadiabatic}. In the study, the connection formula was applied without perturbations, so only the connection formula was considered when crossing the Stokes line extending from the turning point of the upper half plane. Furthermore, there is no discussion of the coefficients included in the connection matrix, which was discussed in Sec.~\ref{subsec:connection_multi}. From this point of view, the discussion in the study is considered partially inadequate.

\begin{figure}[H]
    \centering
    \includegraphics[width=0.7\linewidth]{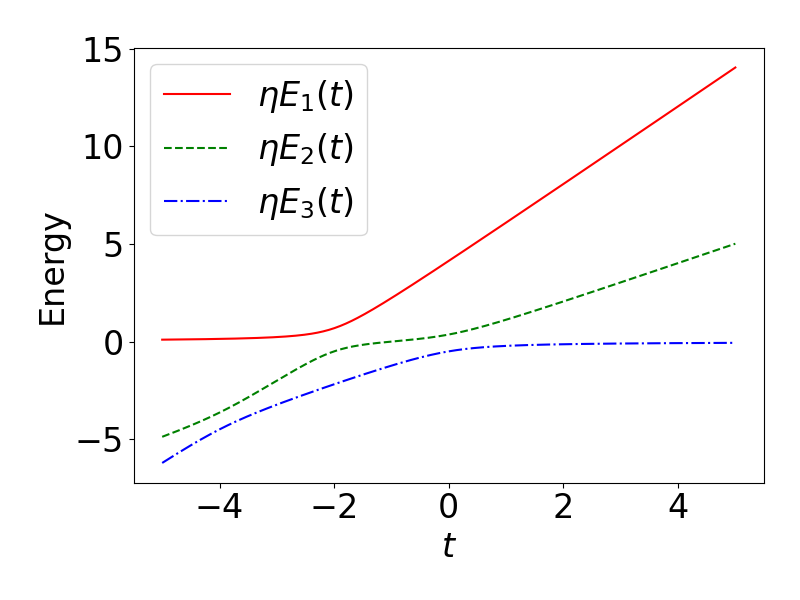}
    \caption{The time dependence of the energy eigenvalues of the Hamiltonian~\eqref{eq:hamiltonian_3level} is plotted. The parameters are set to $\eta v_1=1,\Delta_{12}=\Delta_{13}=\Delta_{23}=\frac{1}{2}\sqrt{v_1/\eta},v_2/v_1=2,a=4\sqrt{v_1/\eta}$. The same parameters are used in all subsequent figures.}
    \label{fig:energy_3level}
\end{figure}

\section{Conclusion\label{sec:conclusion}}

In this paper, we have derived non-perturbative approximations for the adiabatic time evolution of discrete-level quantum systems using the exact WKB analysis. The approximations for the two-level system that we derived are similar to existing known approximations. The key difference is that our derived formula is applicable even when the initial state is not the ground state. Moreover, we derived the approximations for the multilevel system using the exact WKB analysis. Although non-perturbative analysis for multilevel system was mentioned in the previous study~\cite{wilkinson2000nonadiabatic}, our result showed that the dynamics are more complex than those mentioned in~\cite{wilkinson2000nonadiabatic}. Through a concrete example, we have seen that the approximations which we derived agree well with numerical calculations. From this, it can be inferred that the existence of virtual turning points does not generally affect the dynamics.

In this study, we focused on a simple Hamiltonian without singularities, yet Hamiltonians with singularities have also gained attention recently~\cite{cardoso2023landau}. Since the Stokes lines arise from such singularities as well~\cite{koike2000exact}, it is necessary to derive a new connection matrix. It is also known that there exists an integrable model in the time-dependent Hamiltonian~\cite{sinitsyn2018integrable}. By analyzing this dynamics using the exact WKB analysis, one may be able to investigate the conditions for integrable models. These issues will be the subject of future work.

\section*{Acknowledgement}
We thank H. Nakazato and H. Taya for helpful discussions.

\appendix

\section{The connection between global WKB solutions and local WKB solutions in two-level system}\label{appsec:connection}
\subsection{Return to Airy equation near a turning point}
From now on, we consider the behavior near a turning point $t=t_c$. We assume that three Stokes lines extend from this turning point to infinity. The Hamiltonian in the adiabatic basis, such as $H_{A}(t,\eta)$ in \eqref{eq:adi_sch_2_abbreviation}, exhibits singular behavior near the turning point, making it difficult to analyze. Therefore, the basis is transformed in order to perform the analysis in the standard basis. Using the matrix
\begin{align}
    S(t):=\mqty(e^{-\frac{i}{2} \varphi(t)}\cos\frac{\theta(t)}{2}& e^{-\frac{i}{2} \varphi(t)}\sin \frac{\theta(t)}{2}\\ e^{\frac{i}{2} \varphi(t)}\sin\frac{\theta(t)}{2}&-e^{\frac{i}{2} \varphi(t)}\cos \frac{\theta(t)}{2}),
\end{align}
we get
\begin{align}
    \sum_{i=1}^2 a_i(t,\eta)\ket{E_i(t)}
    &=S(t)\ket{\psi_A(t,\eta)}.
\end{align}
It is evident that the state, when expanded in the adiabatic basis $\ket{E_i(t)}$ and the standard basis $\ket{i}$, is related through the matrix $S(t)$. We refer to the state in the standard basis as $\ket{\tilde\psi_C(t,\eta)}$, defined by
\begin{align}
    \ket{\tilde{\psi}_C(t,\eta)}:=\sum_{i=1}^2 a_i(t,\eta)\ket{E_i(t)}.
\end{align}
We also call the two-level Hamiltonian for this state as $\tilde H_C(t,\eta)$, which satisfy
\begin{align}
    i\frac{\partial}{\partial t}\ket{\tilde{\psi}_C(t,\eta)}&=\tilde H_C(t,\eta)\ket{\tilde{\psi}_C(t,\eta)}.
\end{align}
We note that $H(t,\eta)=\tilde H_C(t,\eta)$ for a two-level system, but they are not equal for a multilevel system (see Sec.~\ref{subsec:multi_to_two_ham}). The Hamiltonian $\tilde H_C(t,\eta)$ can be written as
\begin{align}
    \tilde H_C(t,\eta)&=\eta\boldsymbol{d}(t)\cdot \boldsymbol{\sigma}+\eta \omega(t)I,
\end{align}
in general. To eliminate $\omega(t)$, we define
\begin{align}
    \ket{ \psi_C(t,\eta)}:=e^{i\eta\int_{t_c}^t  \omega(s)ds}\ket{\tilde \psi_C(t,\eta)},
\end{align}
and the equation satisfied by $\ket{\psi_C(t,\eta)}$ is
\begin{align}
    i\frac{\partial}{\partial t}\ket{ \psi_C(t,\eta)}&=\eta\boldsymbol{d}(t)\cdot \boldsymbol{\sigma}\ket{\psi_C(t,\eta)}.
\end{align}

Now that we have obtained the Hamiltonian in the standard basis, we next consider attributing the Schrodinger equation to the Airy equation by focusing on the vicinity of the turning point. Using the matrix
\begin{align}
    T(t,\eta)&=T_0(t)+\sum_{j=1}^\infty\eta^{-j}T_j(t),\\
    T_0(t)&=\mqty(\frac{\sqrt{d_x^2(t)+d_y^2(t)}}{\boldsymbol{d}^2(t)}&-\frac{d_z(t)}{\boldsymbol{d}^2(t)}\\ 0&1),
\end{align}
and
\begin{align}
    F(t)&=\mqty(1&0\\ 0&h(t)),\\
    h(t)&=\frac{-i}{\dot x(t)},\\
    x(t)&=\qty(\frac{-3i}{2}\int^t_{t_c}\sqrt{\boldsymbol{d}^2(s)} ds)^{2/3},\label{eq:def_x}
\end{align}
we define the new state
\begin{align}
    \ket{\phi(x(t),\eta)}:=F(t)T(t,\eta)\ket{\psi_C(t,\eta)}.
\end{align}
This state obeys the Airy equation
\begin{align}
    \frac{\partial}{\partial x}\ket{\phi(x,\eta)}&=\eta\mqty(0&1\\ x&0)\ket{\phi(x,\eta)}.
\end{align}

\subsection{Stokes phenomenon in the Airy equation}
The WKB solutions of the Airy equation can be expressed as
\begin{align}
    \ket{\phi_+(x,\eta)}&=\frac{1}{2}\mqty(x^{-1/4}\\ x^{1/4})e^{\eta \int_0^x\sqrt{x'}dx'},\label{eq:phi_p}\\
    \ket{\phi_-(x,\eta)}&=\frac{1}{2}\mqty( x^{-1/4}\\- x^{1/4})e^{-\eta \int_0^x\sqrt{x'}dx'}.\label{eq:phi_m}
\end{align}
We call these the local WKB solutions. By superposition of these WKB solutions, the solution of the Airy equation is expressed asymptotically as
\begin{align}
    \ket{\phi(x,\eta)}\simeq \sum_{j=\pm}c_j\ket{\phi_j(x,\eta)}.
\end{align}
It is known, however, that the coefficient of superposition $c_j$ changes discretely when the solutions are connected across the Stokes line. This is called the Stokes phenomenon, and the Stokes phenomenon for the Airy equation has been well studied.
First, when the solutions are connected counterclockwise across the Stokes line where $\ket{\phi_+(x,\eta)}$ is dominant, the solution varies as in
\begin{widetext}
\begin{align}
    \ket{\phi(x,\eta)}&\simeq c_+\ket{\phi_+(x,\eta)}+c_-\ket{\phi_-(x,\eta)}\\
    &\to c_+\ket{\phi_+(x,\eta)}+(c_-+ic_+)\ket{\phi_-(x,\eta)}.
\end{align}
On the other hand, when the solutions are connected counterclockwise across the Stokes line where $\ket{\phi_-(x,\eta)}$ is dominant, the solution varies as in
\begin{align}
    \ket{\phi(x,\eta)}&\simeq c_+\ket{\phi_+(x,\eta)}+c_-\ket{\phi_-(x,\eta)}\\
    &\to (c_++ic_-)\ket{\phi_+(x,\eta)}+c_-\ket{\phi_-(x,\eta)}.
\end{align}
If we connect the solutions clockwise, the sign is reversed in both cases.

\subsection{Connection of local and global WKB solutions}
Now that the connection formulas for the WKB solutions of the Airy equation are known, the connection formulas for the global WKB solution can be derived by relating the global WKB solutions to the local WKB solutions of the Airy equation.
By Eq.~\eqref{eq:def_x}, the local WKB solutions \eqref{eq:phi_p}, \eqref{eq:phi_m} can be transformed as
\begin{align}
    \ket{\phi_+(t,\eta)}
    &= \frac{1}{2}\mqty(\qty(\frac{-3i}{2}\int^t_{t_c}\sqrt{\boldsymbol{d}^2(s)} ds)^{-1/6}\\\qty(\frac{-3i}{2}\int^t_{t_c}\sqrt{\boldsymbol{d}^2(s)} ds)^{1/6})e^{-i\eta\int_{t_c}^t\sqrt{\boldsymbol{d}^2(s)} ds},\\
    \ket{\phi_-(t,\eta)}
    &= \frac{1}{2}\mqty(\qty(\frac{-3i}{2}\int^t_{t_c}\sqrt{\boldsymbol{d}^2(s)} ds)^{-1/6}\\-\qty(\frac{-3i}{2}\int^t_{t_c}\sqrt{\boldsymbol{d}^2(s)} ds)^{1/6})e^{i\eta\int_{t_c}^t\sqrt{\boldsymbol{d}^2(s)} ds}.
\end{align}
Then, we get the relation of local and global WKB solutions
\begin{align}
    &e^{-i\eta\int_{t_c}^t  \omega(s)ds}S^{-1}(t)T^{-1}(t,\eta)F^{-1}(t)\ket{\phi_+(t,\eta)}\\
    &=-\qty(\frac{-3i}{2}\int^t_{t_c}\sqrt{\boldsymbol{d}^2(s)} ds)^{-1/6}e^{i\int_{t_c}^tg_{11}(s) ds}e^{\frac{i}{2}\varphi(t)}\cos\frac{\theta(t)}{2}\ket{\psi_{A,1}(t,t_c,\eta)},\label{eq:connect_p}
\end{align}
and 
\begin{align}
    &e^{-i\eta\int_{t_c}^t  \omega(s)ds}S^{-1}(t)T^{-1}(t,\eta)F^{-1}(t)\ket{\phi_-(t,\eta)}\\
    &=-\qty(\frac{-3i}{2}\int^t_{t_c}\sqrt{\boldsymbol{d}^2(s)} ds)^{-1/6}e^{i\int_{t_c}^tg_{22}(s) ds}e^{\frac{i}{2} \varphi(t)}\sin\frac{\theta(t)}{2}\ket{\psi_{A,2}(t,t_c,\eta)},\label{eq:connect_m}
\end{align}
where we use the definitions \eqref{eq:psi_p} and \eqref{eq:psi_m}. From these relation, when the solutions are connected counterclockwise across the Stokes line where $\ket{\psi_{A,1}(t,t_c,\eta)}$ is dominant, the WKB solutions varies as in
\begin{align}
    \ket{\psi_{A,1}(t,t_c,\eta)}&\to \ket{\psi_{A,1}(t,t_c,\eta)}+ i\tan\frac{\theta(t_c)}{2}\ket{\psi_{A,2}(t,t_c,\eta)}.\label{eq:connection_tc_p}
\end{align}
On the other hand, when the solutions are connected counterclockwise across the Stokes line where $\ket{\psi_{A,2}(t,t_c,\eta)}$ is dominant, the solution varies as in
\begin{align}
    \ket{\psi_{A,2}(t,t_c,\eta)}&\to \ket{\psi_{A,2}(t,t_c,\eta)}+i\cot\frac{\theta(t_c)}{2
    }\ket{\psi_{A,1}(t,t_c,\eta)}.\label{eq:connection_tc_m}
\end{align}
If we connect the solutions clockwise, the sign is reversed in both cases.

\end{widetext}

\section{Relation between our results and previous research}\label{appsec:coeff}

\subsection{Previous research}\label{appsec:pre_research}
In this section, we assume that $t_c$ is the turning point in the upper half plane unless otherwise specified.
\subsubsection{DDP formula}
In the original derivation of the DDP formula, a two-level system satisfying $d_y(t)=0$ was considered. This derivation was extended to the general two-level systems~\cite{kitamura2020nonreciprocal}. In the following, we present the sketch of the derivation according to~\cite{kitamura2020nonreciprocal}.

They assume that 
\begin{align}
    \boldsymbol{d}^2(t) &\simeq i \alpha\left(t-t_c\right),\label{eq:assump1}\\
    \dot{\boldsymbol{d}}^2(t) &\simeq \beta,\label{eq:assump2}\\
    (\boldsymbol{d}(t) \times\dot{\boldsymbol{d}}(t)) \cdot\ddot{\boldsymbol{d}}(t) &\simeq \gamma\label{eq:assump3}
\end{align}
hold in the vicinity of the turning point $t_c$.
We expand the state as 
\begin{align}
    |\psi(t)\rangle&=\sum_{n=1}^2 c_n(t) e^{-i \int_{t_0}^t \left(\eta E_n\left(s\right)+ g_{n n}\left(s\right)\right)ds}\ket{E_n(t)}.\label{eq:state_exp_c}
\end{align}
We assume the initial conditions as $c_2(t_0)=1$ and $c_1(t_0)=0$ which mean that the state is the ground state, and we also assume $t_0$ is sufficiently past. Then, we get
\begin{align}
    &i  c_1(t)\\
 &\simeq  \int_{t_0}^t  e^{-i \int_{t_0}^{s} \left(\eta \left(E_2\left(s'\right)-E_1\left(s'\right)\right)+g_{22}\left(s'\right)-g_{11}\left(s'\right)\right)ds' }g_{12}(s) ds\label{eq:first_order_approx}
\end{align}
by a first order approximation. If the integration path is changed along the anti-Stokes line~\footnote{We note that the definition of Stokes line and anti-Stokes line may differ between physics and mathematics.} which is defined as the set of $t$ satisfying
\begin{align}
    \operatorname{Im} \int_{t_c}^t E(s) d s=0,
\end{align}
the time evolution is adiabatic except in the vicinity of the turning point, and the non-adiabatic process can be considered only in the vicinity of the turning point, where the behavior of the Hamiltonian does not depend on the model under the assumptions \eqref{eq:assump1}, \eqref{eq:assump2}, and \eqref{eq:assump3}. From this fact, the transition probability is approximately given by
\begin{align}
    &P_{\mathrm{DDP}}\\
    &=\qty|e^{-i \int_{t_0}^{t_c} \qty(\eta\qty(E_2\left(s\right)-E_1\qty(s))+ g_{22}\left(s\right)-g_{11}\qty(s))ds-i\arg g_{12}(t_c)}|^2.
\end{align}
If there is more than one turning point which we call $t_{c,k}$, it can be extended as in 
\begin{widetext}
\begin{align}
    &P_{\mathrm{DDP}}
    =\qty|\sum_ke^{-i \int_{t_0}^{t_{c,k}} \qty(\eta\qty(E_2\left(s\right)-E_1\qty(s))+ g_{22}\left(s\right)-g_{11}\qty(s))d s-i\arg g_{12}(t_{c,k})}|^2.\label{eq:ddp}
\end{align}
\end{widetext}

We note that this method~\cite{kitamura2020nonreciprocal} and another method proposed in~\cite{joye1991interferences} are shown to give the same result in~\cite{kitamura2020nonreciprocal}, although the method~\cite{kitamura2020nonreciprocal} is better in the sense that it yields the gauge invariant formula.

\subsubsection{GDDP formula}
We assume a two-level system satisfying $d_y(t)=0$. If there are multiple turning points $t_{c,k}$ in the upper half plane, then according to the GDDP formula~\cite{suominen1992parabolic,vitanov1999nonlinear}, the transition probability in the adiabatic limit is approximately
\begin{align}
P_{\mathrm{GDDP}}&=\left|\sum_{k=1}^N \Gamma_k e^{-i \eta \int_0^{t_{c,k}}\qty(E_2(s)-E_1(s)) d s}\right|^2, \label{eq:gddp}\\
\Gamma_k&=4  \lim _{t \rightarrow t_{c,k}}\left(t-t_{c,k}\right) g_{12}(t).
\end{align}
We note that there is no rigorous derivation of this formula although this formula is often used~\cite{suominen1992parabolic,vitanov1999nonlinear,guerin2002optimization,lehto2012superparabolic,ashhab2022nonlinear}.

\subsection{Relation to our results}\label{appsubsec:coeff}

In this section, we mention the relation between the connection matrices~\eqref{eq:connect_matrix_2},~\eqref{eq:connect_matrix_1} and the DDP formula~\eqref{eq:ddp} and the GDDP formula~\eqref{eq:gddp}.

First, we consider the GDDP formula~\eqref{eq:gddp}. When $d_y(t)=0$,
\begin{align}
    d_{z}(t_{c,k})&= \pm i d_{x}(t_{c,k})
\end{align}
holds at the turning point $t=t_{c,k}$, so
\begin{align}
    4\lim_{t\to t_{c,k}} (t-t_{c,k})g_{12}(t)
    &=\pm 1  
\end{align}
holds.
On the other hand, regarding the coefficients contained in the connection matrices~\eqref{eq:connect_matrix_2} and \eqref{eq:connect_matrix_1}, 
\begin{align}
    \tan\frac{\theta(t_{c,k})}{2}&=\tan\qty(\frac{1}{2}\arctan \frac{d_x(t_{c,k})}{d_z(t_{c,k})})\\
    &=\mp i
\end{align}
holds. It follows that state changes as 
\begin{align}
    &\ket{\psi_{A,2}(t,t_0,\eta)}\\
    &\to \ket{\psi_{A,2}(t,t_0,\eta)}\mp e^{-i\int_{t_0}^{t_{c,k}} \eta \qty(E_{2}(s)-E_1(s))ds} \ket{\psi_{A,1}(t,t_0,\eta)}
\end{align}
when the state crosses counterclockwise the Stokes line where $\ket{\psi_{A,2}(t,t_{c,k},\eta)}$ is dominant. Considering that this happens every time the state crosses the Stokes line extending from multiple turning points, it can be seen that the GDDP formula~\eqref{eq:gddp} is reproduced.

Next, consider the relation with the DDP formula~\eqref{eq:ddp}. Let $d_y(t)\neq 0$. Let $t$ approach $t_{c,k}$ from the direction of $\zeta$. In this case, the coefficients of the DDP formula~\eqref{eq:ddp} are obtained as 
\begin{align}
    \arg g_{12}(t_{c,k})&=\arg \qty(\pm (t-t_{c,k}))=\zeta \pm \frac{\pi}{2}-\frac{\pi}{2},
\end{align}
and the coefficients of the DDP formula also match our derived formula.

The only difference between our theory and previous theories is the restriction of the initial state. In the DDP formula and the GDDP formula, the initial state is assumed to be the ground state. On the other hand, our theory gives the time-evolution matrix which is approximately unitary. The difference results from whether the turning points in the lower half plane are considered. Next, we will show the role of these turning points with an example.

\bibliography{ref} 
\end{document}